\documentclass[aoas,MSNbibl,nameyear,dvips]{arximspdf}
\usepackage{graphicx}
\doi{10.1214/13-AOAS678} 
\volume{8}
\issue{1}
\pubyear{2014}
\firstpage{232}
\lastpage{258}

\begin{document}
\begin{frontmatter}

\title{Using informative priors in the estimation of mixtures over time
with application to aerosol particle size distributions}
\runtitle{Using informative priors in the estimation of mixtures}

\begin{aug}
\author[a]{\fnms{Darren} \snm{Wraith}\corref{}\thanksref{t1}\ead[label=e1]{dwraith@unimelb.edu.au}},
\author[b]{\fnms{Kerrie} \snm{Mengersen}\ead[label=e2]{k.mengersen@qut.edu.au}\thanksref{t2}},
\author[b]{\fnms{Clair} \snm{Alston}\ead[label=e3]{c.alston@qut.edu.au}\thanksref{t2}},
\author[c]{\fnms{Judith}~\snm{Rousseau}\ead[label=e4]{rousseau@ceremade.dauphine.fr}\ead[label=u1,url]{http://www.foo.com}\thanksref{t3}}
\and
\author[d]{\fnms{Tareq} \snm{Hussein}\ead[label=e5]{tareq.hussein@helsinki.fi}\thanksref{t4,m1}}
\runauthor{D. Wraith et al.}
\affiliation{Institut National de Recherche en Informatique et en
Automatique (INRIA)\thanksmark{t1},
Queensland University of Technology\thanksmark{t2},
CREST, INSEE and Universit{\'e} de Paris Dauphine\thanksmark{t3}, and
University of Helsinki and
University of Jordan\thanksmark{t4}}
\address[a]{D. Wraith\\
INRIA\\
Laboratoire Jean Kuntzman\\
Mistis Team \\
655 Avenue de l'Europe\\
Montbonnot 38334 Saint-Ismier Cedex\\
France\\
\printead{e1}}
\address[b]{K. Mengersen\\
C. Alston\\
School of Mathematical Sciences\\
Queensland University of Technology\\
Brisbane\\
Australia \\
\printead{e2}\\
\phantom{E-mail:\ }\printead*{e3}}
\address[c]{J. Rousseau\\
CREST, INSEE\\
15 Boulevard Gabriel Peri\\
Malakoff Cedex 1 92245\\
France\\
and\\
Universit{\'e} de Paris Dauphine\\
Place du Mar\'{e}chal de Lattre de Tassigny \\
75775 Paris Cedex 16\\
France\\
\printead{e4}}
\address[d]{T. Hussein\\
University of Helsinki\\
Department of Physics\\
Division of Atmospheric Sciences\hspace*{23.5pt}\\
PO Box 48, FI-00014 UHEL\\
Helsinki\\
Finland \\
and
University of Jordan\\
Faculty of Science\\
Department of Physics\\
Amman 11942\\
Jordan\\
\printead{e5}}
\end{aug}
\thankstext{m1}{Supported by the Academy of Finland through the Finnish Center of Excellence (project number
1118615).}

\received{\smonth{3} \syear{2012}}
\revised{\smonth{8} \syear{2013}}

%
\begin{abstract}
The issue of using informative priors for estimation of mixtures at
multiple time points is examined. Several different informative priors
and an independent prior are compared using samples of actual and
simulated aerosol particle size distribution (PSD) data. Measurements
of aerosol PSDs refer to the concentration of aerosol particles in
terms of their size, which is typically multimodal in nature and
collected at frequent time intervals. The use of informative priors is
found to better identify component parameters at each time point and
more clearly establish patterns in the parameters over time. Some
caveats to this finding are discussed.
\end{abstract}

%

\end{frontmatter}

\section{Introduction}

Aerosol particles have a direct and indirect impact on the earth's
climate. One of the most important physical properties of aerosol
particles is their size, and the concentration of particles in terms of
their size is referred to as the particle size distribution. An
important characteristic of these data is that because aerosol
particles are governed by formation and transformation processes they
tend to form well distinguished modal features. Investigating these
features provides an understanding of the dynamic behaviour of aerosol
particles, their effect on the climate and their association with
adverse health effects. This type of data is increasingly being
measured on a regular basis, with the potential to provide more
detailed information than, for example, measurements of particle mass
concentrations such as PM10 or PM2.5, traditionally used for regulatory
purposes [\citet{who06}].

The most common approach for representing particle size distributions
is by treating the size distribution at any time point as a set of
individual typically normal distributions or modes
[\citet{hussein05}, \citet{whitbyetal91}]. In this
formulation the estimation of particle size distributions is then
analagous to a finite parametric mixture model problem at each time
point.

While interest is in the representation of the particle size
distribution as a mixture at each time point, it is also of interest to
describe how this distribution evolves over time. To better understand
aerosol dynamic processes, a feature of the measurements of particle
size distributions is that they are often collected at regular points
in time, and often at quite small time intervals (e.g., every 10
minutes). In this setting, parameters of the mixture model at each
time point are likely to be correlated with neighboring time points and
useful information about the parameters may be gained by incorporating
this information in estimation.

The standard setting in which mixture models have been applied has
largely been for independent random samples [\citet{marinetal05}],
but literature is developing for situations in which the data are
spatially and/or temporally structured [\citet{fernandezgreen02},
\citet{greenrichardson02}, \citet{alstonetal07},
\citet{dunson06}, \citet{caronetal07}, \citet{ji10}].
The development has largely been driven by the increasing availability
of information in a wide variety of applications. An example includes
analysing images (CAT scan) of sheep over time in which interest is in
changes to the composition of fat, bone and tissue
[\citet{alstonetal07}]. In \citet{ji10} interest is in cell
fluorescent imaging tracking modelled using a dynamic spatial point
process and a mixture representation for the different intensity
functions observed.

For the air pollution example considered in this paper, we are
interested in a~mixture representation using a missing data approach,
in which the components themselves can be interpreted as potential
substrata of the data and for which further interest is in their
behaviour over time. In particular, we are interested in the evolution
of the parameters of the components over time, which in the case of the
particle size distribution data is able to reveal important information
about the number and change in size of particles for particular modes
along with a measure of their variation. This information can then be
used to better understand the potential variables affecting the dynamic
behaviour of each mode (e.g., from local effects such as combustion
from petrol and diesel vehicle engines, construction activity, wind
speed, temperature, etc., and regional effects), which are likely to
vary substantially between modes, and provide for a more accurate risk
assessment of potential effects on adverse health outcomes (e.g.,
respiratory illnesses).

Popular recent approaches that allow for the correlated nature of the
parameters in a mixture setting, both within and across epochs, include
Dependent Dirichlet Process mixture models (DDPM) and (spatial) dynamic
factor models (SDFM) [\citet{maceachern99}, \citet{dunson06},
\citet{caronetal07}, \citet{ji10},
\citet{stricklandetal10}]. While these approaches are appealing in
the context of our case study, they do have some drawbacks.
Importantly, interpretation of component parameters is less
straightforward under the DDPM, and the SDFM typically requires
relatively long time series [\citet{stricklandetal10}]. An
alternative that we consider here is the use of informative priors at
each epoch in a finite parametric mixture setting, where the
information required at each epoch is obtained from neighbouring
epochs. This has appeal both in terms of a general Bayesian learning
framework and in terms of interpretability of the mixture components
and weights, which is important in our application. Moreover, while
some of the methods developed for mixture models in the spatial setting
[e.g., \citet{fernandezgreen02}] can potentially be adapted for
use in a time series setting, the influence or choice of informative
priors in a time series framework and the implications in different
data environments have largely not been examined.

In this paper we explore three different informative priors for
estimation of mixtures where the data are highly correlated, and all
parameters in the mixture are allowed to vary. Different simulated data
sets, with features similar to actual particle size distribution data,
are used to highlight the influence of using informative priors and to
identify situations where placing informative priors may not be
beneficial.

The paper is structured as follows. In Section~\ref{secpsddata} we
briefly describe particle size distributions and provide an
illustration with actual data. In Section~\ref{secmixture} we outline
the finite parametric mixture model setup for a single time point and
then outline the three approaches to estimation of a mixture model at
multiple time points. Section~\ref{secres} presents results on the
performance of the approaches on several simulated data sets and actual
data, and we conclude in Section~\ref{secdiscussion} with some
discussion and possibilities for further work.

%
\begin{figure}[t]
\includegraphics{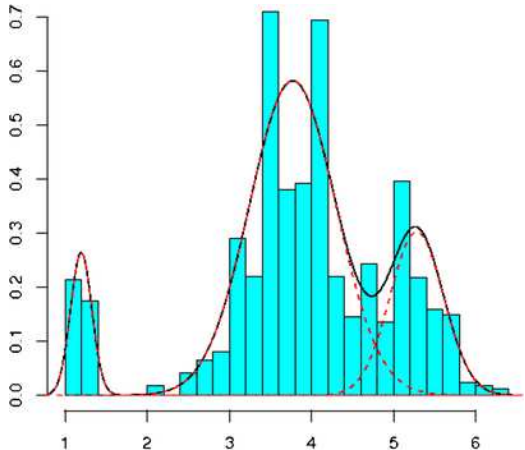}
\caption{Histogram (normalised) of particle size data for
a single time point [$x$-axis: $\log(D_p\mbox{ (nm)}$)]. The black (overall) and
red (components) lines show the inferred density from estimation using \mbox{RJMCMC}.}\label{figplotexamplefit}
\end{figure}

\section{Particle size distribution data}\label{secpsddata}

Figure~\ref{figplotexamplefit} shows an example of particle size
distribution data for one measurement or time point. 
The histogram shows the number of particles $N$ per cubic centimeter
binned by particle size, with the horizontal axis representing the
natural logarithm of the particle diameter in nm [$\log(D_p)$]. The
histogram is normalised, so that its total area equals 1.

%
\begin{figure}[t]
\includegraphics{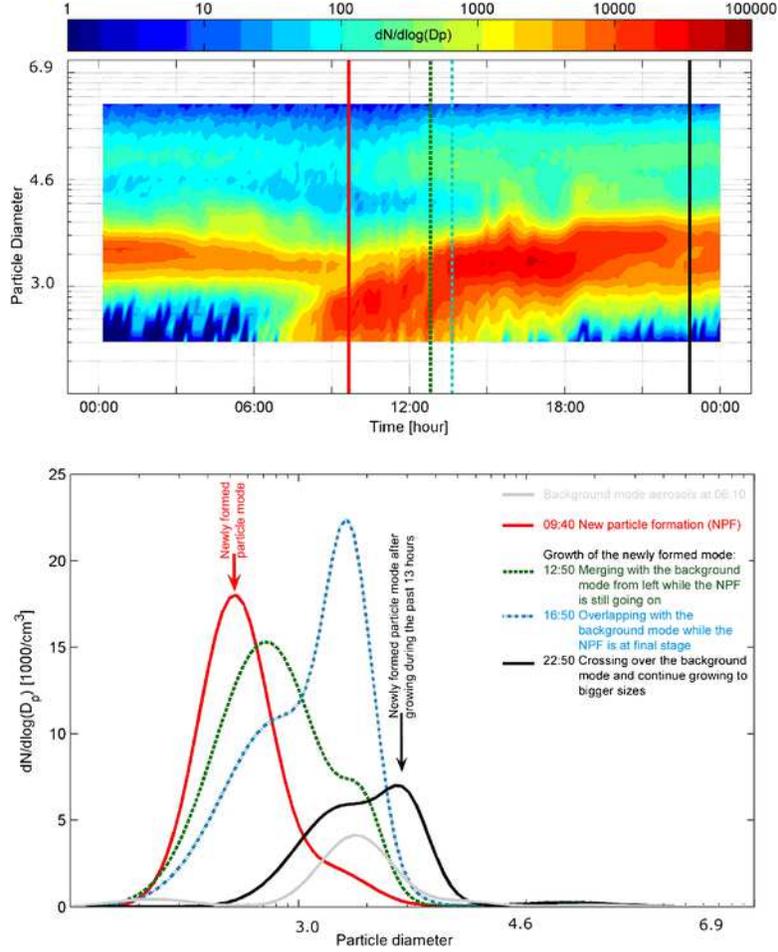}
\caption{An illustration of a new particle formation
event at a Boreal Forest site located in Southern Finland.
(Top)~the temporal variation of the particle size distribution and
(bottom)~selected particle size distributions showing the different stages of
the newly formed particle mode. Particle diameter in (top) $y$-axis and
(bottom) $x$-axis is on the $\log$ scale [i.e., $\log(D_p\mbox{ (nm)})$].}\label{figplotoneday}
\end{figure}

Because aerosol particles are charged, their size can be determined
from their electrical mobility [\citet{mcmurry00}] and a common
instrument that utilises this principle is the Differential Mobility
Particle Sizer (DMPS) [\citet{aaltoetal01}].

In this study we present, as an example, the aerosol particle evolution
before, during and after a new particle formation event at a Boreal
Forest in Southern Finland (Figure~\ref{figplotoneday}). This data set
was selected as it provides a wide ranging representation of modes for
particle size distributions [\citet{dalmasoetal05}]. Because
aerosol particles are governed by formation and transformation
processes, they tend to form well distinguishable modal features. For
example, during background conditions in the Boreal Forest the particle
number size distribution of fine aerosols [diameter $<$2500 nm ($\log$,
7.82)] is bimodal: an Aitken mode [below 100~nm ($\log$,~4.60)] and an
accumulation mode (over 100~nm). During a new particle formation event
a new particle mode, which is commonly known as a nucleation mode, is
formed in the atmosphere with geometric mean diameter below 25 nm
($\log$, 3.22). However, in the urban atmosphere aerosol particles are
more dynamic because of the different types and properties of sources
of aerosol particles and may show more than three modes. Typically the
number concentrations of aerosol particles in the urban background can
be as high as $5\times10^{4}$~cm$^{-3}$ and close to a major road they
often exceed $10^{5}$~cm$^{-3}$ [\citet{kulmalaetal04}].

\section{Methods}\label{secmixture}
In this section we briefly describe the mixture model, outline a~two
stage approach to estimation of parameters over time, and describe
three types of priors for temporal evolution of the parameters.

\subsection{Mixture representation}

The density of data ($y$) at a given time period may be represented by
a finite parametric mixture model
%
%
\begin{equation}
\label{eqnmix} p(y|\theta) = \sum_{j=1}^{k}
\lambda_{j}f(y|\theta_{j}),
\end{equation}
where $k$ is the number of components\vspace*{-1pt} in the mixture,
$\lambda_j$ represents the probability of membership of the $j$th
component ($\sum_{j=1}^{k}\lambda_{j}=1$), and $f(y| \theta_{j}$) is
the density function of component $j$ which has parameters
$\theta_{j}$.

Let $i=1,\ldots,N$ indicate the observed data index. As component
membership of the data is unknown, a computationally convenient method
of estimation for mixture models is to use a hidden allocation process
and introduce a latent indicator variable $z_{i}$, which is used along
the lines of a missing variable approach to allocate observations
$y_{i}$ to each component [\citet{marinetal05}].

In this paper we adopt the common assumption of fitting normal
distributions to aerosol particle size distribution data
[\citet{whitbymcmurry97}]. As PSD data are often measured with a
definite lower and upper bound for the size of the particles, we
introduce a slight modification and assume that the data follow
a~truncated normal distribution. As is commonly assumed, we take the
data ($y$) to be the $\log$ of particle diameters (nm), and the
parameters ($\theta_{j}$) for each component are the mean ($\mu_j$) and
variance ($\sigma_j^{2}$). The number of components $k$ is also assumed
to be unknown.

In the first stage of the temporal analysis, for each time period we
implemented a reversible jump Markov chain Monte Carlo (RJMCMC)
algorithm [\citet{richardsongreen97}]. Although this algorithm is
easily fit at a single time point, the use of RJMCMC for mixture models
with temporal data requires significant preprocessing with respect to
mixing coverage and convergence, as well as postprocessing to provide
adequate summary statistics and between time component mapping.

As an alternative, we considered a two-stage approach. In the first
stage, the number of components was estimated at each time point using
RJMCMC. In the second stage, we fixed the number of components ($k$) to
the maximum observed at any time point and independently estimated the
parameters of the mixture model ($\mu$, $\sigma$ and $\lambda$) for
each time point using a MCMC sampler algorithm. Details of the MCMC
scheme used for the different cases are given below. As we do not
observe all of the components in every time point, we allow component
weights to be ``effectively'' zero [$\inf(\lambda_{t})=0.001$] if
required [for details on the asymptotic behaviour of the posterior
distribution using this approach see \citet{rousseaumengersen10}].
Estimation of parameters of the components that are effectively
``empty'' under this criterion will then essentially be governed by
their respective prior information. For the results to follow in
Section~\ref{secres} we thus only plot the parameters of components
which are not ``empty''.

Priors for the first stage of the analysis were
\begin{eqnarray*}
p(\mu_j) & = & \mathcal{N} \bigl(\xi,\kappa^{-1} \bigr),
\\
p \bigl(\sigma_j^{2} \bigr) & = & \operatorname{IG}(\delta,\beta),
\\
p(\beta) & = & \operatorname{Gamma}(g,h),
\\
p(\lambda) & = & \operatorname{Dirichlet}(\alpha_1,\ldots,\alpha_k),
\\
p(k) & = & \operatorname{Uniform}(1,10),
\end{eqnarray*}
where $\xi$, $\kappa$, $\delta$, $\alpha$, $\eta$, $g$, $h$ are
hyperparameters.

For the second stage, priors were
%
%
\begin{eqnarray}
p(\lambda) & =& \operatorname{Dirichlet}(\alpha _{1},\ldots, \alpha_{k}),\nonumber
\\
p \bigl(\mu_{j}|\sigma^{2}_{j} \bigr) & =&
\mathcal{N} \biggl(\xi_{j},\frac{\sigma^{2}_{j}}{n_{j}} \biggr),\label{eqnprior}
\\
p \bigl(\sigma^{2}_{j} \bigr) & =& \operatorname{IG} \biggl(\frac{v_{j}}{2},\frac{s^{2}_{j}}{2} \biggr),\nonumber
\end{eqnarray}
where again $\alpha_{j},\xi_{j},n_{j},v_{j}$ and $s_{j}$ are
hyperparameters, detailed below. The prior for $\mu$ and $\sigma^2$
could alternatively be decoupled and expressed as in stage~1, but we
did not see a noticeable difference in the results
(Section~\ref{secres}) using either form. For the independent prior
case, we use uninformative priors for $\mu$, $\sigma$ and $\lambda$.


\subsection{Choice of temporal prior}\label{sectempprior}

In the second stage, four priors were considered for linking parameter
values ($\mu,\sigma,\lambda$) over time. The first of these was the
independent prior, in which the correlated nature of the data was
ignored completely and parameters were independently estimated at each
time point. The second, third and fourth were termed the ``informed
prior'', ``penalised prior'' and ``hierarchical informed prior'', as
described below.

\subsubsection{Informed prior}\label{secip}

In this approach we use the information provided from the previous time
period as prior information for the current period. For the main
results we focus on a simple case where posterior estimates from the
previous period are used as prior information for the current period.
We do this to illustrate the influence of a simple prior specification
on the posterior estimates of parameters~($\theta$).

In the case of a mixture model using Gaussian distributions, we have
three parameters ($\mu$, $\sigma$ and $\lambda$) for which we could
utilise available prior information to aid in estimation. Preliminary
investigation indicates that all three parameters are likely to show
strong evidence of autocorrelation, so here we examine the effect of
smoothing on each of these parameters.

For $p(\lambda)$, we allow $\alpha_{j}$ in equation~(\ref{eqnprior}) to
reflect prior information about $\lambda_{j,t-1}$. Thus, we set
$\alpha_{j}=\theta_{j}\bar{m}_{j,t-1}$, where $\bar{m}_{j,t-1}$ is the
mean of the number of observations allocated to component $j$ in the
previous time period, and $\theta_{j}$ is fixed at some value. An
alternative is to impose a distribution on $\theta$, say, $\theta_{j}
\sim\mathrm{U} (0,1)$ [or $\mathcal{N}(1,0.5$)], but we do not present
the results for this approach in this paper.

For the specification of prior information for $\mu$ and $\sigma$, we
set $\xi_{jt}=\mu_{j,t-1}$, $v_{j}=n_{j}/\sigma_{j,t-1}^{2}$ and
$s_{j}=n_{j}$ [to ensure $p(\sigma^2_{jt})$ is centred on
$\sigma^2_{j,t-1}$] and increase the value of $n_{j}$ from the value
set for the independent case to reflect the degree of dependency for
these parameters from the previous period.

\subsubsection{Penalised prior}\label{secpp}

In this approach we base the priors at time $t$ on the aggregated
information at all other time periods. This can be achieved by
employing a reparameterisation of the prior to reflect the degree of
dependency between parameters. \citet{gustafsonwalker03} proposed
a prior for $\lambda$ (in a different context) which can be used in our
setting to downweight large changes in probabilities in successive time
periods. Let
$\underline{\lambda}=(\lambda_{jt},j=1,\dots,k,t=1,\dots,T)$, then
$p(\underline{\lambda})$ is defined as
%
%
\begin{equation}
p(\underline{\lambda}) \propto\operatorname{Dirichlet}(1,\dots,1) \exp
\Biggl(-\frac{1}{\phi}\sum_{t=2}^{T}\|\underline{
\lambda}_{t}-\underline{\lambda}_{t-1}\|^{2}
\Biggr), \label{eqngustaf}
\end{equation}
where smaller values of $\phi$ imply greater smoothing.

A potential advantage of using information about estimates both
forwards and backwards in time is the additional information this may
provide to guide parameter estimates in the current period. This may be
most useful if large changes in the parameter estimates occur for
single periods of time. For the purposes of comparison with
Section~\ref{secsimulateddata}, we compare the results of using a
similar formulation for $\lambda$ in the informed prior approach
(without smoothing on $\mu$ and $\sigma$).

Thus, prior distributions $p(\mu)$ and $p(\sigma)$ are set as for the
independent approach [equation~(\ref{eqnprior})].

For this formulation, we sampled from the posterior distribution of
$\lambda$ using a~rejection sampling approach outlined in
Appendix~\ref{app1}.

\subsubsection{Hierarchical informed prior}\label{secpp}

In this approach an informative prior is placed at two different
levels. The aim of allowing for different levels is to provide
flexibility to the form in which prior information is given in the
model. This flexibility may be needed in cases where the correlation
structure can vary greatly over time: instead of imposing a smoothing
structure directly on strongly varying parameters, we can provide a
less restrictive smoothing through the hyperparameters.

For the hierarchical approach, we will focus on parameters $\mu$ and
$\lambda$ as they are the main parameters of interest for the PSD data
(see Section~\ref{secrealdata} and the
\hyperref[secdiscussion]{Discussion}). The hierarchical approach for
$\mu$ is specified as
%
%
\begin{eqnarray}\label{eqnsmoothmu}
\mu_{jt} &\sim&\mathcal{N}\bigl(\phi_{jt},\varepsilon^{(d)}_{\mu} \bigr),
\nonumber\\[-9pt]\\[-9pt]
\phi_{jt} & \sim&\mathcal{N} \bigl(\phi_{j,t-1}, \varepsilon^{(s)}_{\phi} \bigr),\nonumber
\end{eqnarray}
where $\varepsilon^{(d)}_{\mu}$ and $\varepsilon^{(s)}_{\phi}$ are
scalars, reflecting the variability of $\mu_{jt}$ and $\phi_{jt}$,
respectively. Under this formulation, $\mu$ is used to estimate the
mixture distribution at the level of the data, and $\phi$ represents
the underlying correlation of $\mu$ over time [assuming in this case an
AR(1) process]. In this setting, we can interpret the ratio
$\varepsilon^{(s)}_{\phi}/\varepsilon^{(d)}_{\mu}$ as reflecting the
amount of information we have about the underlying behaviour (signal)
of $\mu$ in comparison to estimates at the level of the data (noise).

For the first time period ($t=1$), we set $\phi_{jt}=\mu_{jt}$. For
estimation of $\mu$ and $\phi$, we use a Gibbs sampling scheme. For
details see Appendix~\ref{app2}.

For the parameter $\lambda$, the Dirichlet prior used in
equation~(\ref{eqnprior}) for the independent model and the informed
prior approach is a very common prior for discrete probabilities. A
natural extension to the Dirichlet prior with a temporal component is
to use its representation in terms of a Gamma distribution.
However, the inflexibility of the Gamma distribution makes it difficult
to construct a temporal structure to the Dirichlet prior. An
alternative formulation of the Dirichlet in terms of the Beta
distribution does not appear to provide greater flexibility.

Another alternative is to use a Logistic-Normal prior for $\lambda$,
where
%
%
\begin{eqnarray}\label{eqnsmoothweight1}
W_{t} & \sim&\mathcal {N}_{k-1}(X_{t}, \Sigma_{d}),
\nonumber\\[-8pt]\\[-8pt]
\lambda_{jt} & =& \frac{\exp(W_{jt})}{\sum_{j=1}^{k-1}{\exp(W_{jt})}}\nonumber
\end{eqnarray}
and where $X_{t}$ is the mean value (number of particles) at time
period $t$.

Using this functional form, the parameterisation of $\lambda$ in terms
of a multivariate normal distribution allows for a suitably flexible
form in which to explore a hierarchical structure for this parameter.
Such flexibility, in comparison to the Dirichlet distribution, has been
investigated in a hierarchical approach for pooling of estimates across
different sampling units [\citet{hoff03}].

In a hierarchical setting and similar to the model used for $\mu$, we
can further say that
%
%
\begin{eqnarray}\label{eqnsmoothweight2}
X_{t} & \sim&\mathcal{N}_{k-1}(X_{t-1},\Sigma_{s}),
\nonumber\\[-8pt]\\[-8pt]
\gamma_{jt} & =& \frac{\exp(X_{jt})}{\sum_{j=1}^{k-1}{\exp(X_{jt})}},\nonumber
\end{eqnarray}
where $\Sigma_{d}$ and $\Sigma_{s}$ reflect the variability of $W_{t}$
and $X_{t}$, respectively. Analogous to the above discussion for $\mu$,
under this formulation the parameter $\lambda$ is used to estimate the
mixture model at the level of the data, and $\gamma$ represents the
underlying or smoothed behaviour of $\lambda$ over time, which may be
prone to large fluctuations from the data.

For the simulation results and actual data to follow, we specify the
diagonal entries of $\Sigma_{d}$ and $\Sigma_{s}$, and fix off-diagonal
entries to be zero. For comparability with the hierarchical approach
for $\mu$, and using similar notation for the smoothing parameters, we
specify $\Sigma_{d}=\varepsilon^{(d)}_{\lambda}I_d$ and
$\Sigma_{s}=\varepsilon^{(s)}_{\gamma}I_d$. The interpretation of
$\varepsilon^{(d)}$ and $\varepsilon^{(s)}$ is then the same as before,
but this time in terms of $\lambda$.

For estimation of $\lambda$ and $\gamma$ we use a Gibbs sampling scheme
with a Metropolis Hastings step. For details see Appendix~\ref{app2}.
For identifiability both $W_{t}$ and $X_{t}$ are $k-1$ dimensional, and
$\lambda_{k}=1-\sum_{j=1}^{k-1}\lambda_{j}$ (with the\vspace*{-1pt}
same identification used for~$\gamma$).

In practice, it may be difficult to specify \textit{a priori} the
parameter values for $\varepsilon^{(d)}$ and $\varepsilon^{(s)}$, as
little information about the variability of the parameters for the
mixture components may be known. Estimation of these parameters also
requires a choice to be made about the degree of smoothing required.
For\vspace*{1pt} the purposes of this paper we focus on specifying the
parameter values for $\varepsilon^{(d)}$ and $\varepsilon^{(s)}$, and
explore briefly the effect on the results of varying these values. In
the discussion we talk about this issue further. For now, one approach
to specifying $\varepsilon^{(d)}$ may be to use the results from a
RJMCMC approach used in the first stage, estimate or explore the
variability of $\mu$ or $\lambda$ over time,\vspace*{1pt} and then use
this information to set the parameter values for $\varepsilon^{(d)}$.
The parameter value for $\varepsilon^{(s)}$ could be set as a smaller
multiple of~$\varepsilon^{(d)}$, and be varied to assess the influence
of the results. Using the information from the results of the
independent approach, for a fixed upper bound $k$ could also be used.
Although there is extra computational time involved in running either
approach in the first stage, such a strategy may prove useful in order
to assess the influence of different prior information on the results.

\subsection{Labelling issues}

Where component parameters are themselves the subject of the analysis,
an important and commonly encountered issue in Bayesian mixture
modelling relates to the labelling of these parameters during the MCMC
run. As the likelihood of a mixture is by definition multimodal, using
exchangeable or noninformative priors can (and should) result in
parameters moving freely over the parameter and component labelling
space during sampling. (In theory, for a $k$ component mixture, $k!$
permutations of the labelling of the parameters are possible.)
Estimation of functionals of these parameters (conditional on
labelling)\vadjust{\goodbreak} at the end of sampling is thus then
problematic. Several empirical approaches to deal with this issue
(commonly called ``label switching'') have been proposed in the
literature [\citeauthor{stephens97} (\citeyear{stephens97,stephens00}),
\citet{celeuxetal00}, \citeauthor{fruwirth01}
(\citeyear{fruwirth01,fruwirth06}), \citet{jasraetal05},
\citet{marinetal05}, \citet{sperrin10}, \citet{yao11}],
generally by relabelling parameters in proximity to one of the $k!$
modal regions during the run [e.g., \citet{fruwirth01}] or at the
end of sampling [e.g., \citet{stephens97}].

Under exchangeable or noninformative priors the main reason for label
switching relates to a lack of identifiability of the mixture model
(particularly with respect to enforcing a unique labelling). Given two
component parameter sets $\theta_j$ and $\theta_k$, a finite mixture
model is weakly identifiable if at least one element of $\theta_j$ and
$\theta_k$ differs. For practical purposes then, the element that
differs can be used to enforce a unique labelling. For exchangeable
priors, in the sense that priors for $\theta_j$ and $\theta_k$ are the
same, it is clear that there is a lack of identifiability.

In the case of using informative priors on $\theta_j$ and $\theta_k$,
the issue of model identifiability is less clear. Although the use of
different informative priors can help to separate $\theta_j$ and
$\theta_k$, in practice, it will depend on the strength of the
informative prior to separate at least one element of each parameter
set. To our knowledge, there has been very little theoretical
investigation of this issue. In practice, one can assess whether
parameters are well separated by analysing the path of parameters over
the sampling run and/or from plots of marginal densities. However, only
part of the picture may be revealed by doing this. First, a relatively
long sampling run is needed to allow the parameter to fully explore the
space (perhaps many thousands of iterations). In the case of Gibbs
sampling, the sampler may also become trapped in one of the modal
regions [\citet{celeuxetal00}]. Apart from the case of using a
fully exchangeable prior, this second case is more difficult to
identify in practice, as it can be unclear whether the Gibbs sampler is
truly trapped or has found a uniquely labelled parameter space. A
pragmatic solution could be to start the sampler from different values,
although good starting locations can be difficult to determine in
high-dimensional space. Alternative (and often more involved) solutions
could be to: reparameterise and change the conditioning
[\citet{marinetal05}]; use tempering to facilitate more
exploration [\citet{celeuxetal00}]; and/or modify the Gibbs
sampling proposal and acceptance [\citet{celeuxetal00},
\citet{marinetal05}].

For both the independent and informed prior approaches outlined
previously, it is possible to relabel the output using existing
empirical approaches. In these cases, the posterior is updated
sequentially across time and relabelling can take place during or at
the end of each time period. In the results to be presented, we used
the \textit{maximum a posteriori }(MAP) estimate to select one of the
$k!$ modal regions and chose either a distance-based measure on the
space of parameters [\citet{celeuxetal00}], or on the space of
allocation probabilities [\citet{stephens97},
\citet{marinetal05}] to relabel parameters in proximity to this
region.\vadjust{\goodbreak}

In the case of the penalised and hierarchical informed priors, a joint
posterior that includes all components and all parameters over time is
used and sampling updates occur globally over time. Relabelling of
sampling output thus necessitates a permutation of labels for all time
points together. Such a joint approach is similar in spirit to the
Viterbi algorithm approach used in state space models
[\citet{godsilletal01}], but the joint posterior in our case is
different and would require further work outside the scope of this
paper.

Using informative priors that are incompatible with the data can also
force too many distinct components and lead to overfitting, which can
in some cases lead to label switching. The potential for label
switching could arise if two components with similar parameters are
fitted where one component would suffice under the true model. As
before, the similarity of the parameters could result in a lack of
identifiability of the model. In using an informative prior there is
also a model choice issue in terms of selecting the best model for the
data, where ``best'' is defined, for example, as the most parsimonious
model in terms of the smallest value of $k$. In practice,
discrimination between the choice and use of informative priors can be
made by fitting several mixture models to the data. Point-process
representations of the estimated parameters from using a noninformative
prior (such as Figure~\ref{figrjplotmeans}) can also offer a good
visual guide as to the potential range and scope of the parameter
space.


\subsection{Accounting for binning and truncated data}\label{secacctrunc}

Aerosol particle measurements are commonly recorded in the form of a
number of distinct particle size ranges, or channels, the size and
number of the channels being governed by the type and setup of the
measurement instrument. For example, in the sampled data from
Hyyti\"al\"a (see Figure~\ref{figplotexamplefit} and
Section~\ref{secrealdata}), we observed 32 distinct size partitions
(bins) covering the range from 3~nm to 650~nm.

Such coarsening of the data created by binning has an impact on density
estimates and in a mixture context the number of components required to
adequately model the data [\citet{alstonmengersen10}]. To address
this, we add another step in the Gibbs sampler in which we simulate a
new latent variable (say $x$) which is drawn from the believed
underlying density of the data ($y$), in this case the fitted mixture
model at current estimates of the parameters, at each iteration of the
Gibbs sampler. As the sampling takes place within each bin, the
simulation of the latent variable essentially involves sampling from a
truncated Normal distribution for which there are a number of proposed
approaches. For computational efficiency we used the slice-sampling
approach of \citet{robertcasella04}. For details of the approach
and of the Gibbs sampler see \citet{alstonmengersen10}. As the
number of observations within each bin (and hence the latent sample
size) is quite large, an extra step can be added after the latent
variables are simulated in which the samples within each bin are
divided into a number of sub-bins and computations in the Gibbs sampler
proceed based on the new binned data. This can greatly speed up
computations compared to using the full latent sample whilst reducing
the coarsening of the data created by the original bins. In general, we
found\vadjust{\goodbreak} comparable results to the full latent sample
by using 3 additional sub-bins for each original bin (in total 97 bins
are used compared to the original 32 bins).

\section{Results}\label{secres}

In this section we present and assess the results using simulated data
and then present the results of applying the approaches to particle
size distribution data from Hyyti\"al\"a, Finland. We use the simulated
data to test the impact of the different prior representations and the
degree of smoothing. We first use an informative and penalised prior
only on the weights ($\lambda$), and then assess the influence of using
an informative prior on $\mu$ and $\sigma$ in order to assess the
influence of using prior information for each parameter separately.

For the independent, informed and penalised prior approaches the
results are based on 50,000 iterations with a burn-in period of 20,000
(i.e., the first 20,000 samples are discarded). Results using RJMCMC
(used in the case study) are based on 200,000 iterations with a burn-in
period of 100,000. Convergence was assessed by visual inspection and
using the Gelman--Rubin statistic [\citet{brooksgelman98}].

\subsection{Simulated data}\label{secsimulateddata}

\subsubsection{Data setup}
We simulated data sets indicative of the type of behaviour of aerosol
particle size distribution data observed at Hyyti\"al\"a, a Boreal
Forest site in Southern Finland (SMEAR~II) [\citet{vesalaetal98}].
A particular feature of these particle size distribution data is both a
growth in the mean and weight for some of the modes (components) and a
decline in weight for others. Changes can also occur to the variance of
the modes and at times they can follow a similar pattern to the weights
over time.

We simulated data from two different cases. In the first case (D1), we
simulated data which are highly correlated across time, a feature of
particle size distribution data observed in practice for most time
periods where measurements are commonly taken at small time intervals.
This data set was also simulated with parameter estimates where at
times the mixture is not well identified (component means and weights
are not well separated). Of interest in this setting is the effect of
using either the informed prior or penalised prior approach compared to
the independent approach.

In practice, it is quite common to observe sudden large changes in the
number of particles measured which may persist for a number of time
periods. This is more often observed when there are relatively few
particles for a particular size group, and more so for the smaller
sized particles (an example of this type of data is examined in
Section~\ref{secrealdata}). Thus, for the second data set (D2) we
simulated data for the first component where the weight for the smaller
sized particles is quite volatile. For this data set the mixture is
well identified. Further details and results are available in the
supplementary material [\citet{wraithetal13}].

For both cases (D1)~and~(D2), we simulated data using three components
on 32 distinct size partitions (bins) equally spaced (on the $\log$
scale) covering the range\vadjust{\goodbreak} from 3~nm to 20~nm in
particle diameter (on the $\log$ scale 1~to~3). The sample size for
each time period is 1000 and the total number of time periods was 100.
Further details of the sampling process for each case are provided in
Section~\ref{secd1} and the supplementary material
[\citet{wraithetal13}, second data set (D2)].

For the results to follow, except as specified otherwise, for the
independent, informed prior and penalised prior approaches, we set the
hyperparameters to be $\xi=(1.5,3.5,5.0)$; $s^{2}_{t,j} = 10$; $v_{t,j}
= 10/0.6^{2}$; and $n_{j} = 2$, which were chosen to be weakly
uninformative considering the range and size of the data. %
For the independent and informed prior approaches, the original Gibbs
sampling output has been relabelled using a distance-based measure on
the space of parameters [\citet{celeuxetal00}] and also (as a
check) on the allocation space [\citet{stephens97},
\citet{marinetal05}].

\subsection{Simulated data set \textup{(D1)}: Highly correlated data}\label{secd1}

\subsubsection{Smoothing on \texorpdfstring{$\lambda$}{lambda}}

\begin{figure}[t]
\includegraphics{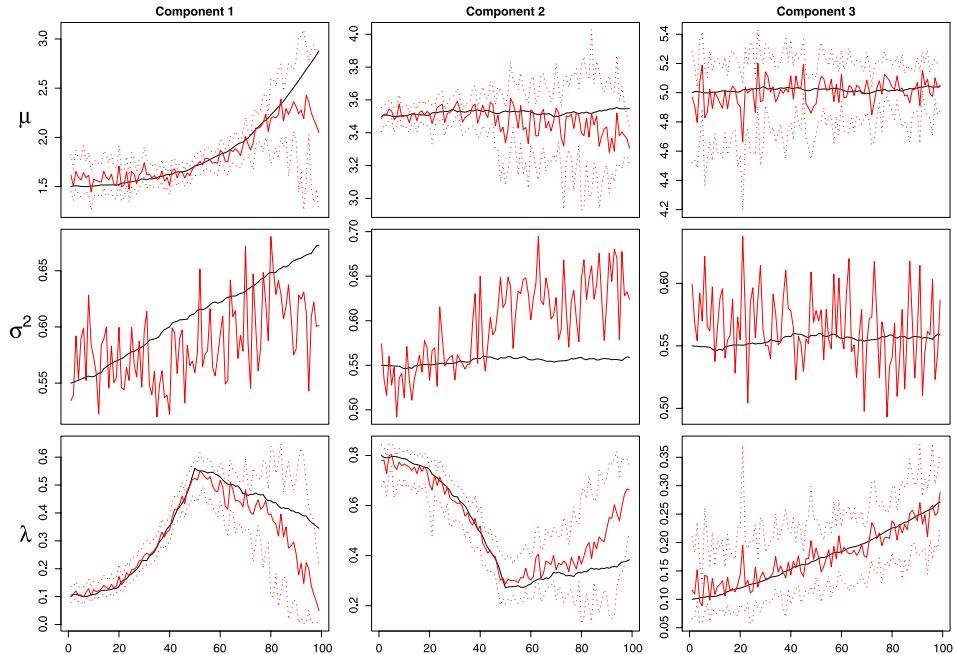}
\caption{Results for simulated data set \textup{(D1)} (highly
correlated data) with growth and influx of new particles for component~1.
Plot of estimated posterior mean for parameters [$\mu$ (top),
$\sigma^2$ (middle) and $\lambda$ (bottom)] over time ($x$-axis) for
independent approach: simulated data (black); independent (red); and
95\% credible interval (dotted line). The columns represent the
components (components 1 to 3).}\label{figsimOactualind}
\end{figure}

As shown in Figure~\ref{figsimOactualind} (black line), for the first
data set \textup{(D1)} we simulated data for the first component with a mean
value increasing from 1.5 to 3.0, and weight increasing from 0.1 to 0.6
and then decreasing to 0.3, over time. Often a consequence of the
growth in the first component is a decline in size and weight for the
larger sized particles and this is reflected in the weight for the
second component following an opposite pattern to the first component.
For the third component, the weight increases from 0.1 to 0.3 over
time. The parameters $\mu$ and $\lambda$ are simulated with some noise
around the parameter values, and the sample size is 1000.

%

Figure~\ref{figsimOactualind} also shows the results of using the
independent approach. We see that at times the parameter estimates for
the independent approach deviate from the actual data.

\begin{figure}[t]
\includegraphics{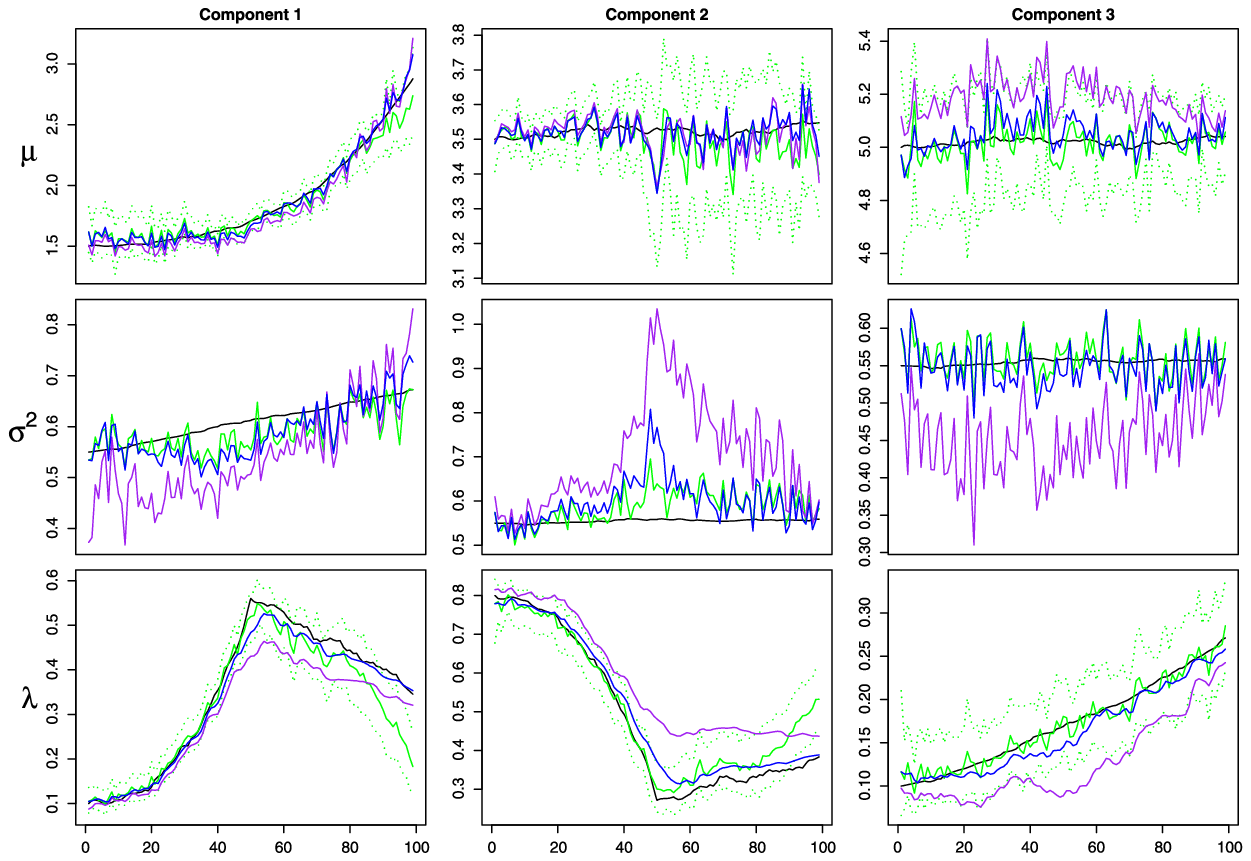}
\caption{Results for simulated data set \textup{(D1)} (highly
correlated data) with growth and influx of new particles for component~1.
Plot of estimated posterior mean for parameters [$\mu$ (top),
$\sigma^2$ (middle) and $\lambda$ (bottom)] over time ($x$-axis) for
informed prior approach: simulated data (black); Theta${}=0.1$ (green);
Theta$={}$0.8 (blue); Theta${}=1.3$ (purple);
and 95\% credible interval (dotted line).
The columns represent the components (components 1 to 3).}\label{figsimOactualIP}
\end{figure}

\begin{figure}
\includegraphics{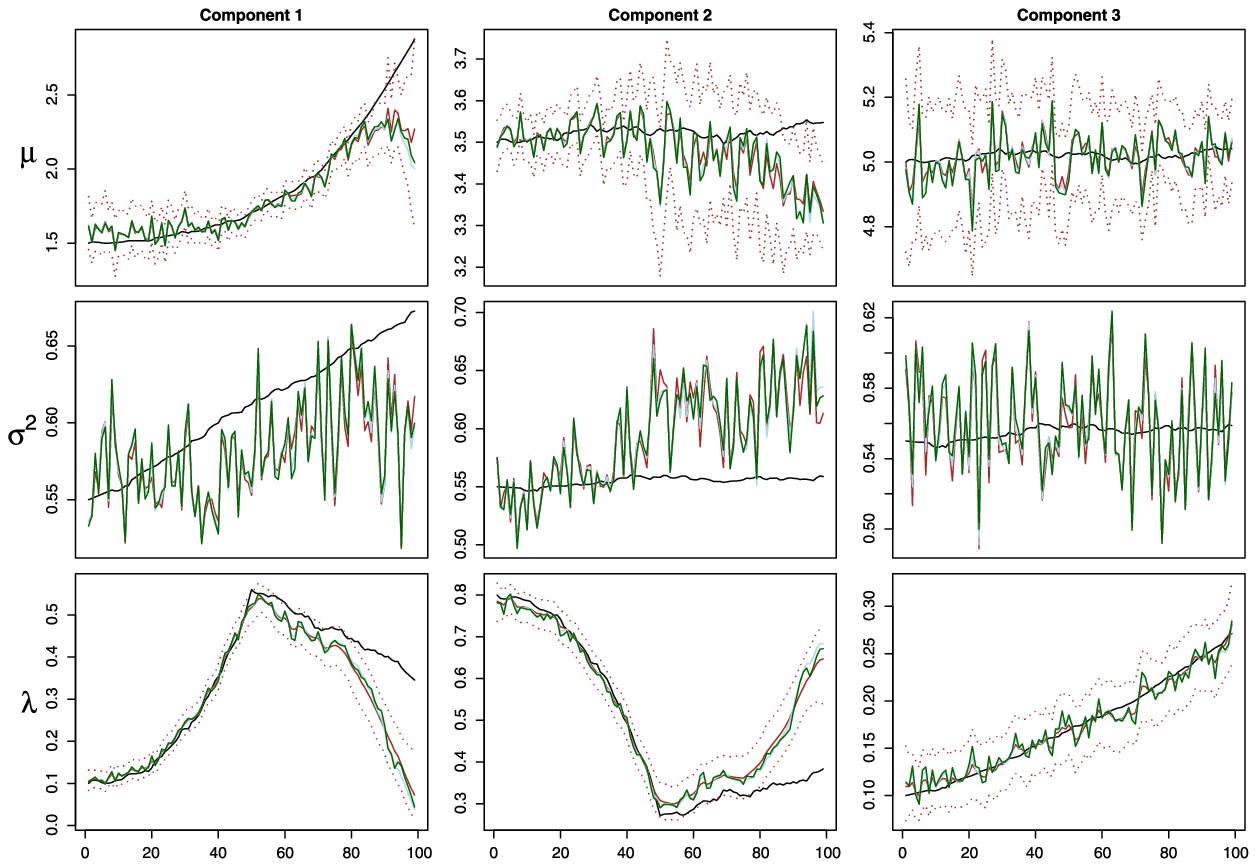}
\caption{Results for simulated data set \textup{(D1)} (highly
correlated data) with growth and influx of new particles for component~1.
Plot of estimated posterior mean for parameters [$\mu$ (top),
$\sigma^2$ (middle) and $\lambda$ (bottom)] over time ($x$-axis) for
penalised prior approach: simulated data (black); $\phi=0.04$ (brown); $\phi
=0.08$ (light blue); $\phi=0.12$ (dark green); and 95\% credible
interval (dotted line). The columns represent the components (components 1 to 3).}\label{figsimOactualPP}
\end{figure}

Figures~\ref{figsimOactualIP}~and~\ref{figsimOactualPP} show the
results for the informed prior and penalised prior compared to the
actual data, respectively. In Figure~\ref{figsimOactualIP}, the results
show the effect of varying the degree of smoothing on $\lambda$ for the
informed prior using $\theta=(0.1,0.8,1.3)$. For the results of the
penalised prior, we vary the degree of smoothing on $\lambda$ using
$\phi=(0.04,0.08,0.12)$.

%
%
%

In Figure~\ref{figsimOactualIP}, we can see that the parameter
estimates for $\lambda$ for all three values of $\theta$ appear to
closely follow the actual data, with the closest estimates to the
actual data being for $\theta=0.8$ and 1.3. As we are only using an
informed prior on the weights, the parameter estimates for $\mu$ and
$\sigma$ appear to be quite variable over time compared to the actual
data. However, the variability appears to be slightly less for these
variables than for the independent approach
(Figure~\ref{figsimOactualind}) and closer to the actual data over
time. Of interest is the closeness of the parameter estimates of $\mu$
and $\sigma$ for components 1 and 2, which more clearly follow the true
growth occurring in component 1 and the stability over time for
component 2 compared to that observed for the independent approach.

In Figure~\ref{figsimOactualPP}, the parameter estimates for the
penalised prior approach appear to deviate slightly from the actual
data for components 1 and 2. For the third component, the parameter
estimates for the penalised prior approach follow the actual data with
some noise. Overall, the results from the penalised prior approach are
similar to the independent approach but with less variability over
time.

\subsubsection{Smoothing on \texorpdfstring{$\mu$}{mu} and \texorpdfstring{$\sigma$}{sigma}}\label{secsensmusigma}

We turn now to an assessment of the impact of using an informative
prior for $\mu$ or $\sigma$ over time. We present results for the
highly correlated data set, since this is the most sensitive of the
simulated data as discussed above. Here we set $n_{j}=25$,
$\xi_{jt}=\mu_{j,t-1}$, $v_{j}=200/\sigma_{j,t-1}^{2}$ and
\mbox{$s_{j}=200$}.

%
%
\begin{figure}

\includegraphics{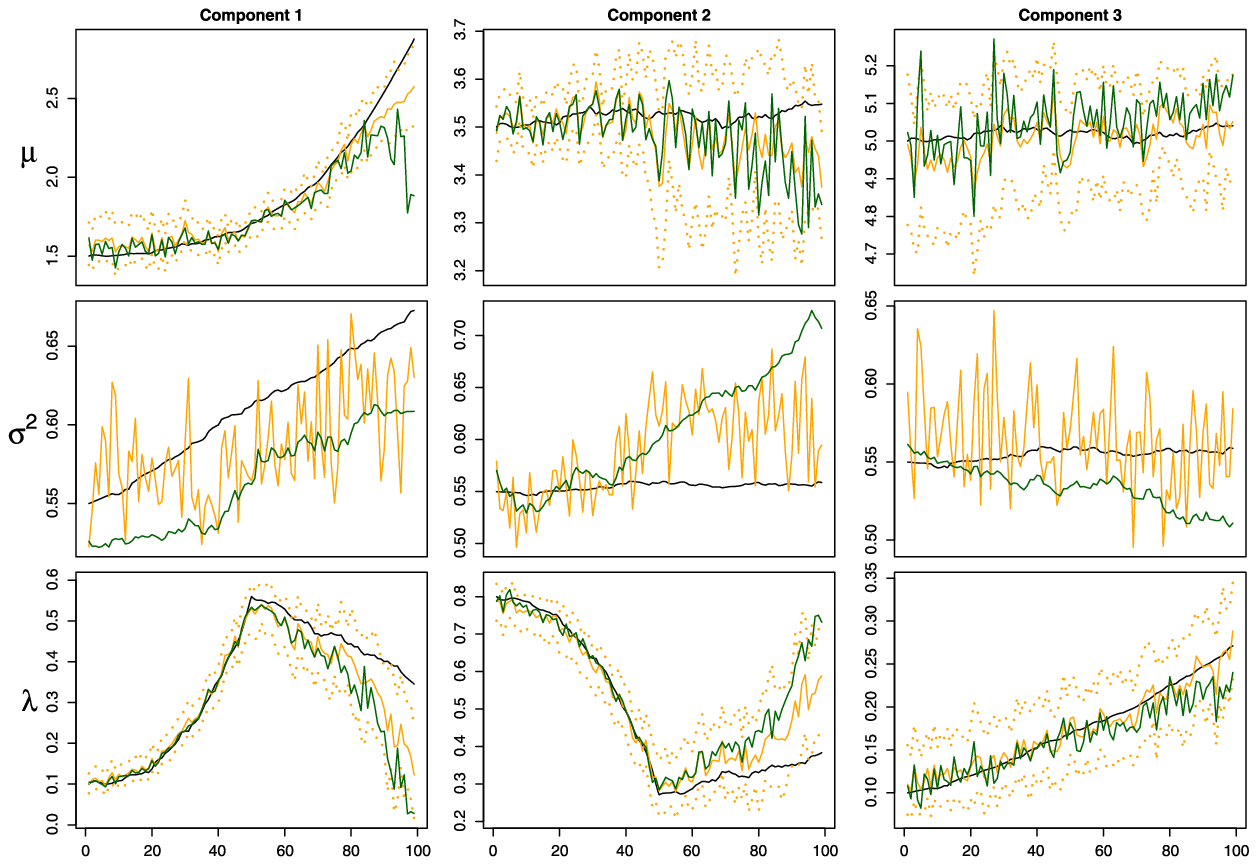}

\caption{Results for simulated data set \textup{(D1)} (highly
correlated data) with growth and influx of new particles for component~1.
Plot of estimated posterior mean for parameters [$\mu$ (top), $\sigma
^2$ (middle) and $\lambda$ (bottom)] over time ($x$-axis) for
informed prior approach: simulated data (black); smoothing on $\mu$ (orange);
smoothing on $\sigma$ (dark green); and 95\% credible interval
(dotted line). The columns represent the components (components 1 to 3).}\label{figsimOactualtpmusigma}
\end{figure}

In Figure~\ref{figsimOactualtpmusigma}, the parameter estimates for the
informative prior for $\mu$ appear to more closely follow the actual
data than using an informative prior for $\sigma$. Although the
parameter estimates for both approaches appear to be further away from
the actual data than using an informative prior for $\lambda$, they do
appear to be closer than under the independent approach.

\subsubsection{Smoothing on \texorpdfstring{$\mu$}{mu} and \texorpdfstring{$\lambda$}{lambda}}\label{secsensmusigma}

\begin{figure}

\includegraphics{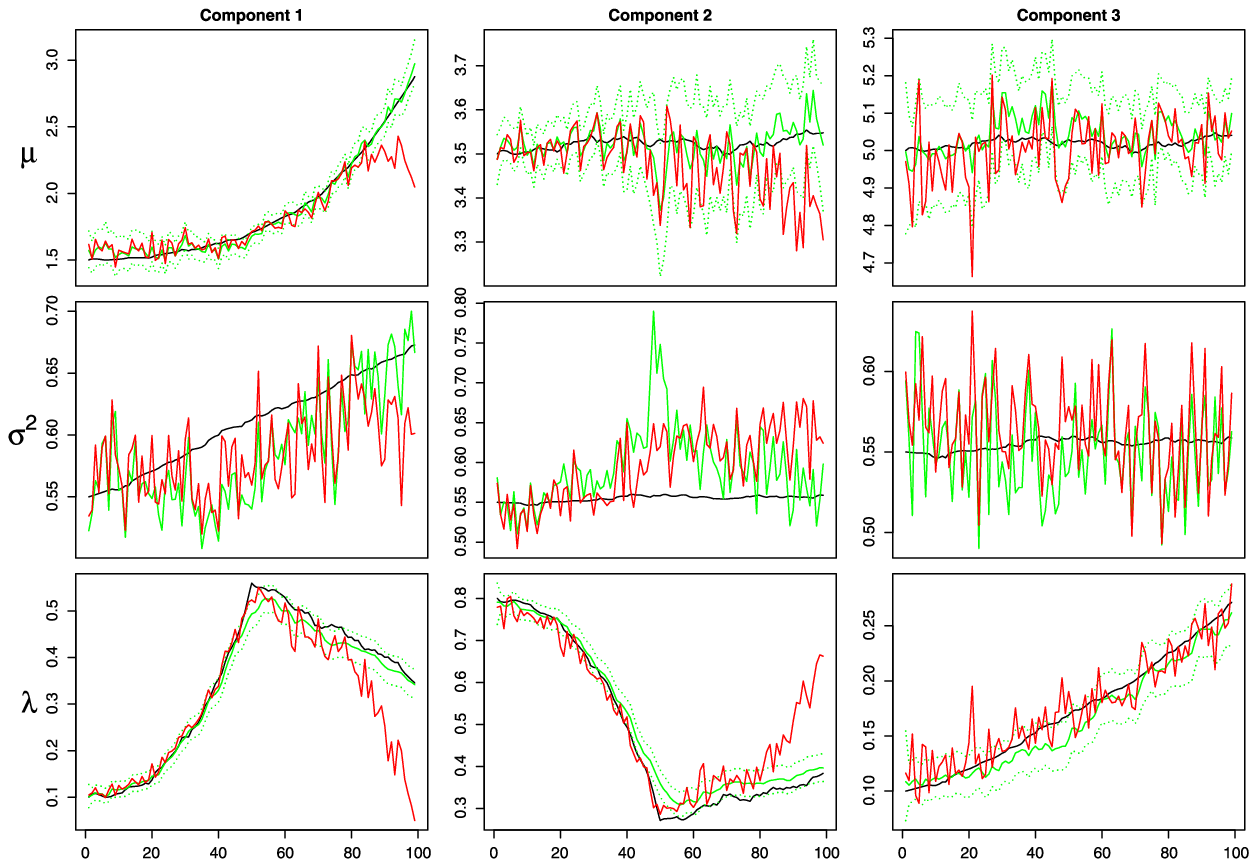}

\caption{Results for simulated data set \textup{(D1)} (highly
correlated data) with growth and influx of new particles for component~1.
Plot of estimated posterior mean for parameters [$\mu$ (top),
$\sigma^2$ (middle) and $\lambda$ (bottom)] over time ($x$-axis) for
approaches: simulated data (black); independent (red); smoothing on
$\mu$ and $\lambda$ (green); and 95\% credible interval (dotted
line). The columns represent the components (components 1 to 3).}\label{figcaseIIsimOsmlambdamu}
\end{figure}

Figure~\ref{figcaseIIsimOsmlambdamu} shows the results of using an
informative prior on both $\mu$ and $\lambda$. In this example, the
results are similar to using an informative prior only on $\lambda$.
Thus, depending on the objectives of the analysis, using an informative
prior on both parameters may not be needed.

%

\subsection{Simulated data set \textup{(D2)}: Noisy data}\label{secd2}

For the second simulated data set, where the weight for the smaller
sized particles is quite volatile, the results of smoothing on $\mu$,
$\sigma$ and $\lambda$ for the informed prior and penalised prior
suggests that large adjustments to one parameter (e.g., from volatility
in some time periods) are not supported unless compensatory measures
can be taken by the other parameters. In contrast to these results, we
do not see any large compensatory adjustments being made to parameters
by using a hierarchically based informative prior for $\gamma$. Details
are available in the supplementary material
[\citet{wraithetal13}].

\subsection{Case study}\label{secrealdata}

The data set studied here was taken from a measurement site at
Hyyti\"al\"a, Finland; a plot of the measurements for the day selected
is shown in Figure~\ref{figplotoneday}. This particular day was
selected as it shows a new particle formation event occurring, whereby
a new mode of aerosol particles appears with a significant influx of
particles (as high as $10^{6}$~per~cm$^{3}$) with a geometric mean
diameter ($<$10 nm), growing later into the Aitken (25--90~nm) or
accumulation modes (100${}+{}$nm). In terms of a temporal mixture model
setting, we will be able to assess the performance of the four prior
specifications outlined previously as new components are introduced and
both a growth in the mean and weight for those components are observed.

The data from Hyyti\"al\"a consists of measurements which were taken
every 10~minutes (144 time periods) and for each time period in the
form of 32 distinct size partitions (bins) equally scaled (on the
$\log$ scale) covering the range of 3~nm to 650~nm (on the $\log$ scale
1 to 6.5).

\begin{figure}

\includegraphics{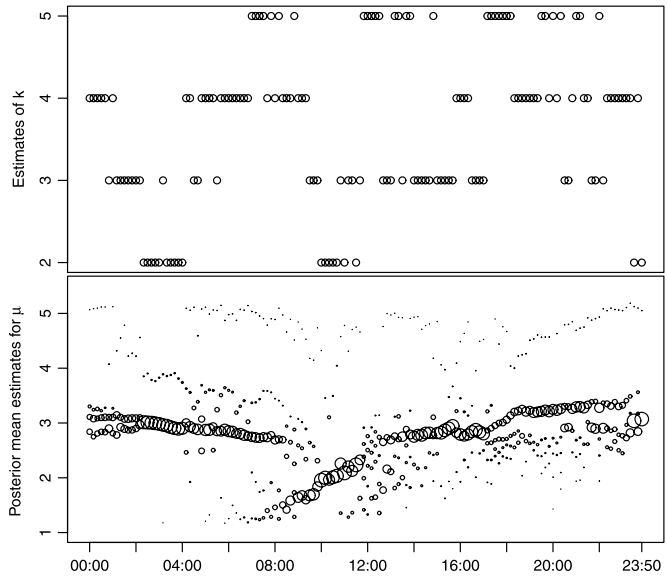}

\caption{(Top)~plot of estimates of $k$ (posterior median) from
RJMCMC algorithm over time (every 10 minutes)
(Hyyti\"al\"a). (Bottom): plot of posterior mean estimates
for $\mu_{j}$ [$\log(D_p\mbox{ (nm)}$)] from \mbox{RJMCMC} algorithm over the same time
period. Stage~1 of analysis for temporal evolution of parameters. The
size of the circles is proportional to the weight ($\lambda_{jt}$)
corresponding to $\mu_{jt}$.} \label{figrjplotmeans}
\end{figure}

As outlined in Section~\ref{secmixture}, the first stage of our
approach is to apply RJMCMC to each time period. These results are then
used to guide the choice of the number of components and initial
parameter estimates for the second stage analysis, in which temporally
correlated priors are used to model the evolution of the mixture
parameters over time. Figure~\ref{figrjplotmeans} shows the results of
the first stage of the algorithm, with a plot of the posterior mean
estimates for $\mu_{jt}$ at each time point $t$ (bottom panel), with
the size of the circles indicating the corresponding weight
$\lambda_{jt}$. The average number of components estimated with the
highest probability over the day was four; the minimum number of
components was one, and the maximum number of components was five (see
top panel of Figure~\ref{figrjplotmeans}).

%

\begin{figure}

\includegraphics{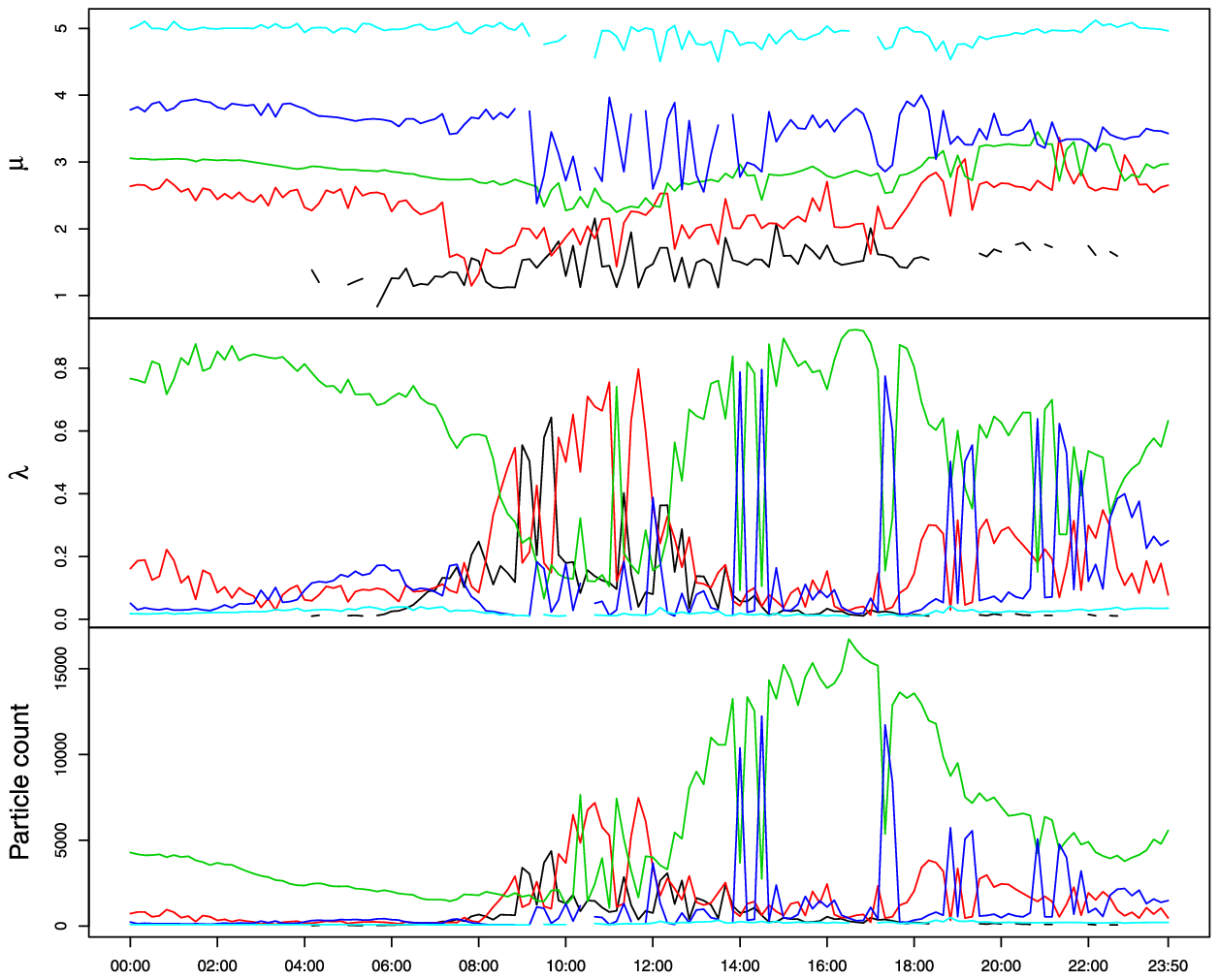}

\caption{Plot of estimated posterior mean for parameters over
time for actual data. Independent approach. Posterior mean estimates
for $\mu$ [(top)~$\log(D_p\mbox{ (nm)}$)], $\lambda$ (middle) and particle count
(per cm$^3$) (bottom). Stage~2 of the analysis for the evolution of
parameters. Measurements taken every 10 minutes. Colours indicate the
components to which parameter estimates belong. (The parameter estimates
for the first component are black, parameters for the second component
are red, for the third component they are green, etc.) Note: only
components where $\lambda> 0.01$ are plotted.}\label{figdm512HMind}
\end{figure}

For the second stage, we fixed the number of components to be five with
the initial mean values equally spaced across the range of possible
diameter values [$\xi=(1.1,2.0,3.0,4.0,5.0)$] and $\kappa=10$.
Figure~\ref{figdm512HMind} shows the results of \mbox{estimation} using
the independent approach (original output has been relabelled using a
distance-based measure on the space of parameters
[\citet{celeuxetal00}]; similar results were obtained by
relabelling on the allocation space [\citet{stephens97},
\citet{marinetal05}]).
%
%

Due to the large number of particles (per cm$^3$) observed over the day
(see Figure~\ref{figplotoneday}), the results of estimation using the
informed prior approach for $\mu$ and/or $\lambda$ are very similar to
the results of the independent approach, so are not shown here. As we
are interested in the effect of using an informative prior in this
context, we rescale the number of particles by a factor of $10$ and
assess the results. The median number of particles over the course of
the day is then 6893, reaching a maximum of 17,740. Using this
rescaling of the data, the results of the independent approach are the
same as shown in Figure~\ref{figdm512HMind}.

\begin{figure}

\includegraphics{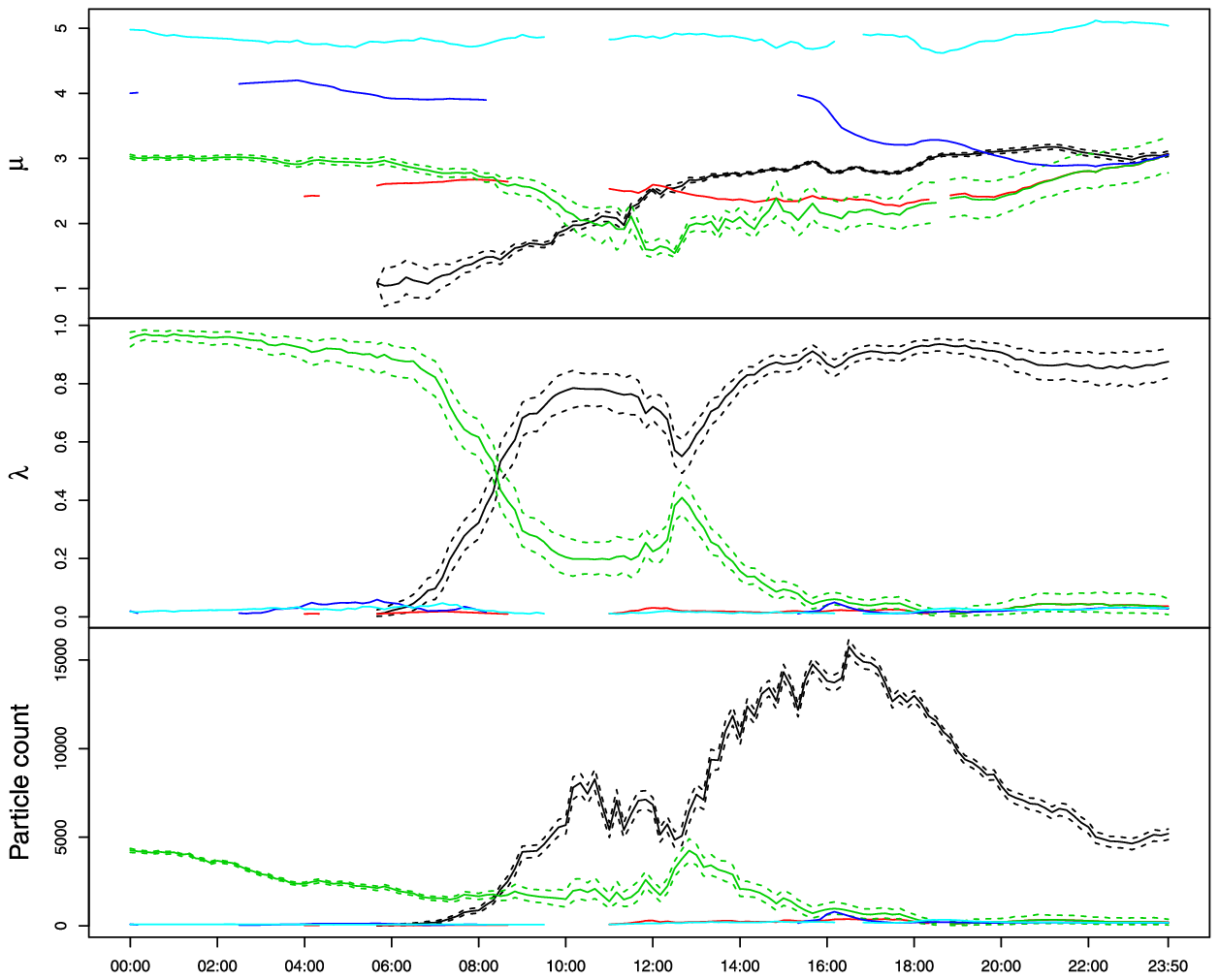}

\caption{Plot of estimated parameters over time for actual data.
Informed prior approach for $\mu$ and $\lambda$. Posterior mean
estimates for $\mu$ [(top)~$\log(D_p\mbox{ (nm)}$)], $\lambda$ (middle) and
particle count (per cm$^3$) (bottom) and 95\% credible interval
(dotted line). Stage~2 of the analysis for the evolution of parameters.
Measurements taken every 10 minutes. Colours indicate the components to
which parameter estimates belong. (The parameter estimates for the first
component are black, parameters for the second component are red, for
the third component they are green, etc.) Note: only components where
$\lambda> 0.01$ are plotted.}\label{figdm512HMlambda}
\end{figure}

Figure~\ref{figdm512HMlambda} shows the results of estimation using the
informed prior approach for $\mu$ and $\lambda$. For these results,
$\theta=0.2$, $n_{j}=15$ and $\xi_{jt}=\mu_{j,t-1}$, which provides for
a moderately informative prior across time. Similar results were
obtained using the hierarchical model with similar strength of prior
information. Of interest to note is that in the original output we did
not see any evidence of label switching within the gibbs sampling runs.
Although there is some evidence of instability between 12:00 and 13:00
for the weight ($\lambda_j$) of two of the components (newly formed and
background particles), this appears to be due to instability in the
data (see Figure~\ref{figrjplotmeans}), with the means of these
components remaining reasonably well separated during this period.

%
%

Compared to the results from the independent approach
(Figure~\ref{figdm512HMind}), the results from the informed prior
approach for $\mu$ and $\lambda$ suggest a clear pattern for both over
the course of the day. In particular, and of interest to aerosol
physicists, is the clear growth pattern shown for the mean of the first
component representing the smaller sized particles. The results
indicate that these particles grow from approximately 1.40 (4~nm) to
3.0 (20~nm) in diameter on the natural $\log$ scale (or in
nanometers).

The path of the parameters for the first component (black) is less
clear using the independent approach, principally as a result of the
newly formed smaller particles merging into the component representing
the background particles (green). 
There is also some evidence of instability in the labelling
\textit{across time}, which is natural given the absence of temporal
information in the prior.


We note that different representations of this data are possible
depending on prior assumptions. If we relax the assumed moderate
correlation of the new and background particles, then there is less
separation between the component associated to the newly formed
particles and to that of the background particles, similar to the
results we see for the independent approach. Ideally, more information
is needed to model these data; this could take the form of greater
assumptions about the underlying stochastic process and the possible
inclusion of external factors (e.g., meteorological data)
impacting/influencing the observed process. We discuss this issue more
in the following section.

\section{Discussion}\label{secdiscussion}

In this paper we explored the problem of estimating Bayesian mixture
models at multiple time points. Under different situations, approaches
that employ information about neighbouring time points compared
favourably to results based on an independent approach. By including
additional temporal information about parameters for correlated time
periods, we may be able to better identify individual components at
each time point. As an aid for inference, we may also be able to obtain
smoother parameter estimates over time and from this be able to clearly
establish patterns or identify anomalies from the data.

The results highlight a number of observations about mixture
representations at multiple time points. First, analysis of the
evolution of parameters of a mixture over multiple time points
highlights the large degree of dependency that exists between component
parameters. Changes to a parameter in one component may flow on to the
parameter in a nearby component. Depending on the context of the
study, we can anticipate this dependency to be more readily apparent
for the weight parameters, but we found similar dependencies to exist
for other parameters. The second is the need to be mindful that the
same parameter in one component may have a different correlation
structure over time to the same parameter in another component. In the
context of particle size distribution data, we often observed greater
volatility in estimates for the smaller particles compared to the
larger sized particles and so at times the correlation structure of the
parameters between these respective components appeared to be quite
different.

A possible effect of using informative priors in this context is to
impose a prior not supported by the data or to impose a temporal
correlation structure where such a structure does not exist, and
thereby cause unnecessary adjustments to other parameters. We observed
this most clearly in the results from the simulated data where at times
the data was quite noisy. For this data set, using an informative prior
for a parameter, which supported large adjustments away from the actual
data, resulted in large compensatory adjustments being made not only by
other parameters within the same component, but also to parameters in
neighbouring components. The easy solution may be to use an appropriate
correlation structure for components, but of course this may not always
be known \textit{a priori}.

A further result of the dependency that can exist between parameters of
components and within component parameters is that the inclusion of
correlation information to aid in the identifiability of the mixture
may not be required for all \mbox{parameters} or, alternatively, all
components. In the context of a mixture with a small number of
components, we may only need to provide more information about one
parameter for an influential component in order to separate out the
influence of competing components. This result will also be useful if
the correlation structure for one parameter or parameters for one
component are more readily known. In the context of a mixture of
Gaussians, we generally found that an informative prior was only needed
on $\mu$ or $\lambda$ or possibly both. This result could well be
context specific and influenced by any reliance on the means for
defining (in terms of size) and ordering of components. The choice of
which parameter to use more information may also be guided by whether
it is a parameter of interest for inference as demonstrated in analysis
of the case study where most interest was in the behaviour of both
$\mu$ and $\lambda$ over time. In this case, and in general, one must
be careful in the analysis of selected parameters, as it can largely be
a conditional analysis in view of the behaviour of other possible
cross-correlated parameters within the same component and between
components.

In the hierarchical informed prior approach the influence of the
informative prior at the two levels was specified by parameters
$\varepsilon^{(d)}$ (low level) and $\varepsilon^{(s)}$ (high level),
and the values assigned to these parameters are critical in carrying
information about the correlation structure of the parameter of
interest. In this paper, we decided to choose parameter values based on
prior belief in the correlation structure of the data; alternatively,
these parameters could be estimated. To this effect, a number of
approaches are available for estimation [\citet{westetal97},
\citet{fahrmeir04}]. However, in order to estimate
$\varepsilon^{(d)}$ and $\varepsilon^{(s)}$, we still face a choice as
to the degree of penalisation or smoothing of the parameter in light of
the apparent variability in the data. This is a common issue in
temporal and spatial modelling in general.

While many of the above difficulties may seem to be avoided if
smoothing approaches are applied retrospectively on parameter estimates
from an independent mixture model, this type of analysis may largely
ignore the true mapping of components or the path of parameters over
time. From the results of the simulated data, the large degree of
dependency that we observe between the parameters of a mixture over
time suggests that including temporal information to better identify
one of the parameters at a single time point can flow on to affect
other parameters. This could change inference about both the mixture
representation at a point in time and also the behaviour of mixture
parameters over time.

In general, one of the potential difficulties in using an informative
prior approach to smooth parameter estimates over time is the variable
degree of influence the prior may have in the posterior. If the primary
objective is to obtain smoothed parameter estimates over time, larger
sample sizes and noisiness of the data at times may warrant
increasingly restrictive priors. In such cases where the objective
might be to downplay the influence of the data, a number of alternative
approaches to increase the influence of prior information can be used
[\citet{ibrahimetal03}]. In all cases, it is valuable to undertake
a sensitivity analysis in order to assess the effect of the prior. Such
an analysis should include the independent prior as a baseline
comparison.

A further limitation of the approach outlined is that it is
computationally expensive. Most of this expense is experienced in the
first stage of the analysis (which can be skipped in the presence of
good prior knowledge of the parameter space). For estimation of PSD
data over one day using 144 time points, the running time of the RJ
approach with 200,000 iterations was approximately 3 hours using an
Intel Centrino 2 processor 2.80~GHz. In comparison, the second stage
approach using 50,000 iterations took approximately one hour. Such
computational expense quickly becomes burdensome if analyses is
required for several days or, indeed, several weeks. Of course, the use
of the first stage for subsequent days may not be required,
considerably reducing the computational time involved.

Although we have focussed on developing a hierarchical approach for
parameters $\mu$ and $\lambda$, we could equally apply the same
approach to consider estimation of~$\sigma$. Such an approach may be to
consider a half-$t$ distribution which has previously been used in
similar hierarchical settings [\citet{gelman06}].

The hierarchical approach considered here can be readily generalised to
include covariates. Moreover, through the flexibility of assuming a
logistic normal distribution on the weights, we can better explore and
estimate transitory movements between components.

In some situations it may be of interest to combine components and
allow components to share a common grouping. This could be of interest
where some components are only needed to account for the skewness of a
larger component or to allow an analysis based on a mixture
representation with fewer components of most interest. Although this
grouping of components could be undertaken retrospectively, it may also
be interesting to see the effects such a grouping has upon estimation
of the parameters and their evolution over time.

For estimation of aerosol particle size distributions, the dynamics of
the aerosol process and the complexity of the influences on particle
concentration and size demand the use of approaches which utilise as
much information from the data as possible. To this end, the inclusion
of temporal information may be helpful.

\begin{appendix}
\section{Penalised prior}\label{app1}
In this section we outline the rejection sampling algorithm for
$\lambda$ proposed by \citet{gustafsonwalker03} for the penalised
prior approach:

Prior
%
%
\begin{equation}\label{eqngustaf}
p(\underline{\lambda}) \propto \operatorname{Dirichlet}(1,\ldots,1) \exp
\Biggl(-\frac{1}{\phi}\sum_{t=2}^{T}\|\underline{\lambda}_{t}-\underline{\lambda}_{t-1}\|^{2}\Biggr).
\end{equation}

Posterior
%
%
\begin{eqnarray}
\qquad && p(\underline{\lambda}|\phi,m)
\nonumber\\[-4pt]\\[-16pt]
&&\qquad \propto\prod_{j=1}^{k}
\Biggl\{ \prod_{t=1}^{T}
\operatorname{Dirichlet}(m_{jt}+1)\mathrm{I}(\lambda_{jt}) \Biggr\}
\exp\Biggl(-\frac{1}{\phi}\sum_{t=2}^{T}
\|\underline{\lambda}_{t}-\underline{\lambda}_{t-1}
\|^{2} \Biggr).\nonumber
\end{eqnarray}

\citet{gustafsonwalker03} suggest sampling $\lambda_{jt}/s$ from a
$\operatorname{Beta}(m_{jt}+1$, $m_{kt}+1)$ distribution and accepting
when $U\leq g_{1}(\lambda_{jt})/g_{2}(\lambda_{jt})$ ($U \sim
\mathrm{U}(0,1)$), where
%
%
\begin{eqnarray}\label{eqnppg1}
g_{1}(\lambda_{jt}) &=& \lambda_{jt}^{m_{jt}}(s-
\lambda_{jt})^{m_{kt}}\mathrm{I}(\lambda_{t})\nonumber
\\
&&{}\times \exp\bigl[-\phi^{-2}\bigl\{(\lambda_{jt}-
\lambda_{j,t-1})^{2}+ \bigl(\lambda_{jt}-(s-
\lambda_{j,t-1}) \bigr)^{2}
\\
&&\hspace*{62pt}{} +(\lambda_{jt}-\lambda_{j,t+1})^{2} + \bigl(
\lambda_{jt}-(s-\lambda_{k,t+1}) \bigr)^{2} \bigr]\bigr)\nonumber
\end{eqnarray}
and
%
%
\begin{eqnarray}\label{eqnppg2}
g_{2}(\lambda_{jt})& =& \lambda_{jt}^{m_{jt}}(s-
\lambda_{jt})^{m_{kt}}\mathrm{I}(\lambda_{t})\nonumber
\\
&&{}\times \exp\bigl[-\phi^{-2}\bigl\{ \bigl(\lambda^{*}-
\lambda_{j,t-1} \bigr)^{2}+ \bigl(\lambda^{*}-(s-
\lambda_{j,t-1}) \bigr)^{2}
\\
&&\hspace*{62pt}{} + \bigl(\lambda^{*}-\lambda_{j,t+1} \bigr)^{2} +
\bigl(\lambda^{*}-(s-\lambda_{k,t+1}) \bigr)^{2} \bigr]\bigr),\nonumber
\end{eqnarray}
where
%
%
\begin{equation}\label{eqnpplambdastar}
\quad \lambda^{*}=\max \bigl\{0,\min  \bigl\{
\tfrac{1}{4}(\lambda_{j,t-1}+s-\lambda_{k,t-1}+
\lambda_{j,t+1}+s-\lambda_{k,t+1}),s \bigr\}  \bigr\},
\end{equation}
$s=\lambda_{jt}+\lambda_{kt}$ and $g_{1}(\lambda_{jt})\leq
g_{2}(\lambda_{jt})$. Here $I(\lambda_{t})$ is an indicator function
equal to~1 when $\lambda_{t}\in[0,1]^2$ and 0 otherwise.

\section{Details of MH Gibbs sampler for hierarchical~model}\label{app2}

\textit{Hierarchical model for $\mu$}

Update $z$, $\beta$, $\lambda$, $\sigma^2$ as in the independent
approach.

Update $\phi$ and $\mu$ by sampling from the conditionals,
\begin{eqnarray*}
\phi_{jt}|\cdot & \sim&\mathcal{N} \biggl(\frac{\varepsilon^{(d)}\phi
_{j,t-1}+\varepsilon^{(s)}\mu_{jt}}{\varepsilon^{(d)}+\varepsilon^{(s)}},
\frac{1}{(\varepsilon^{(d)})^{-1}+(\varepsilon^{(s)})^{-1}} \biggr),
\\
\mu_{jt}|\cdot & \sim&\mathcal{N} \biggl(\frac{\phi_{jt}+m_{j}\bar
{y_{j}}\varepsilon^{(d)}\sigma_{jt}^{-2}}{\varepsilon
^{(d)}m_{j}\sigma_{jt}^{-2}+1},
\frac{\varepsilon^{(d)}}{(\varepsilon^{(d)}m_{j}\sigma
_{jt}^{-2}+1)} \biggr).
\end{eqnarray*}\vspace*{-15pt}\

\textit{Hierarchical model for $\lambda$}

Update $z$, $\beta$, $\mu$, $\sigma^2$ as in the independent approach.
Update $\gamma_{t}$ by sampling from the conditional,
\begin{eqnarray*}
X_{t} &\sim& \mathcal{N}_{k-1} \biggl(\frac{\Sigma
_{d}^{-1}W_{t}+\Sigma
_{s}^{-1}X_{t-1}}{\Sigma_{d}^{-1}+\Sigma_{s}^{-1}},
\frac{1}{\Sigma_{d}^{-1}+\Sigma_{s}^{-1}} \biggr),
\\
\gamma_{jt} &= & \frac{\exp(X_{jt})}{\sum
_{j=1}^{k-1}{\exp(X_{jt})}},
\end{eqnarray*}
where $X_{kt}=0$.

Update $\lambda_{t}$ using a Metropolis Hastings step.

Sample\vspace*{-1pt} from $W_{t} \sim \mathcal{N}_{k-1}
(X_{t},\sigma^{2}_{p}I )$, where $ \lambda_{jt} =
\exp(W_{jt})/\sum_{j=1}^{k-1}\exp(W_{jt})$ and $\sigma^{2}_{p}$ is the
variance of the proposal.

Let $W_{k1}=0$ and for $t=2$, $W_{j2}=\log
(\overline{m}_{j1}/\overline{m}_{k1} )$, where $\overline{m}_{j1}$ is
the mean number of observations allocated to component $j$ in the
previous time period (under the independent approach).
\end{appendix}

\section*{Acknowledgements}
The authors gratefully acknowledge funding from the Australian Research
Council (ARC) as part of two ARC Discovery Projects and the ARC Centre
of Excellence in Complex Dynamic Systems and Control.  The authors are
also grateful for helpful discussions with Christian P. Robert and Tony
Pettitt in early stages of this work, and to comments from the Editor,
Associate Editor and referees.

\begin{supplement}[id=suppB]
\stitle{Simulated data set (D2): Noisy data}
\slink[doi]{10.1214/13-AOAS678SUPP}
\sdatatype{.pdf}
\sfilename{aoas678\_supp.pdf}
\sdescription{Details and results for the second simulated data set (D2) where the data is quite noisy.}
\end{supplement}


%
%

\printaddresses

\end{document}